\documentclass[12pt]{iopart}

\usepackage[utf8]{inputenc}
\usepackage[T1]{fontenc}
\usepackage[english]{babel}
\usepackage[normalem]{ulem}
\usepackage{mathrsfs}

\usepackage{graphicx}
\usepackage{xcolor}

\usepackage{iopams}
\usepackage{amstext}
\usepackage{amssymb}

\usepackage[colorlinks=true, allcolors=blue]{hyperref}
\usepackage[all]{hypcap}  

\newcommand{\bra}[1]{\left\langle #1 \right|}
\newcommand{\ket}[1]{\left| #1 \right\rangle}

\renewcommand{\vec}{\mathbf}

\usepackage[colorinlistoftodos]{todonotes}
\usepackage[colorlinks=true, allcolors=blue]{hyperref}

\makeatletter
\def\subsubsection{\@startsection{subsubsection}{3}{10pt}{-1.25ex plus -1ex minus -.1ex}{0ex plus 0ex}{\normalsize\bf}}
\def\paragraph{\@startsection{paragraph}{4}{10pt}{-1.25ex plus -1ex minus -.1ex}{0ex plus 0ex}{\normalsize\textit}}
\renewcommand\@biblabel[1]{#1}
\renewcommand\@makefntext[1]%
{\noindent\makebox[0pt][r]{\@thefnmark\,}#1}
\DeclareRobustCommand\onlinecite{\@onlinecite}
\def\@onlinecite#1{\begingroup\let\@cite\NAT@citenum\citealp{#1}\endgroup}
\def\tagform@#1{\maketag@@@{\ignorespaces#1\unskip\@@italiccorr}}
\let\orgtheequation\theequation
\def\theequation{(\orgtheequation)}
\makeatother

\renewcommand{\tablename}{\footnotesize Table}

\begin{document}

\title{Correlated many-body noise and emergent $1/f$ behavior in an anharmonic fluctuator model
}
\author{P. N. Thomas Lloyd$^{1,2}$, Valentin Walther$^{2,3}$, and H. R. Sadeghpour$^2$}
\ead{ptlloyd@berkeley.edu}%
\address{$^1$Dept. of Physics, University of California, Santa Barbara, California, 93106, USA}
\address{$^2$ITAMP, Center for Astrophysics | Harvard \& Smithsonian, Cambridge, Massachusetts 02138, USA} 
\address{$^3$Department of Physics, Harvard University, Cambridge, Massachusetts 02138, USA}

\date{}
\maketitle 
\begin{abstract}
Fluctuating electric fields emanating from surfaces are a primary source of decoherence in trapped ion qubits. Here, we show that superradiant phonon-induced excitation exchange between adatoms can lead to a reduction of electric field noise at low temperatures. We derive an exact mapping between the noise spectrum of $N$ fluctuators with $M$ vibrational levels to $N+M-1 \choose N$-1 two-level dipoles. We provide conditions for which the ubiquitous $1/f$ noise can emerge, even though the system is composed of only a single type of fluctuator, thus suggesting a new mechanism for the phenomenon.

\end{abstract}

\section{Introduction}
Prerequisite to all functional quantum information processing protocols is maintaining the quantum coherence and mitigating deleterious noise effects \cite{arute2019,paladino2014}. Electric field noise emanating from surfaces is a particularly worrisome aspect for a host of miniaturized applications in physics. Patch potentials (spatially variant electrostatic potentials on surfaces)  affect force measurements of quantum electrodynamics or non-contact friction at short distances \cite{lamoreaux1997,stipe2001,speake2001}. Surface electric field fluctuation is one of the largest contributor of noise at mHz frequencies for operation of space-based gravitational wave detectors \cite{pollack2008}. 
Electric field fluctuations produced by surface charges on diamond can become a larger source of noise than magnetic noise for near-surface nitrogen vacancy qubits \cite{kim2015,romach2015,myers2017}. 

Experiments \cite{hite2012,hite2017} have shown that surface-induced anomalous low-frequency noise is the major source of decoherence in trapped ion qubit operations.  
Various groups have explored the evolution of this anomalous noise in ion traps at cryogenic, room and elevated temperatures. No single behavior has been observed over the full temperature range \cite{MIT2008,LincolnLab2014,Daniilidis2014,brownnutt2015,berlinudi2021} though noise tends to decrease at lower temperatures. The effect of contaminant density and identity on the surface noise has also been studied with Auger electron spectroscopy of gold electrodes, upon ion bombardment \cite{kim2017,berlinudi2021}. In one experiment the noise power density initially increases as contaminant layers are removed and then decreases with increasing contaminant removal \cite{Sedlacek2018-multi-mechanisms}. Effects of layer covering on ion trap electrode or diamond surfaces have been investigated \cite{safavi-naini2013,chrostoski2018}. Dielectric fluctuations in thin-film covered silicon cantilever tips have been shown to produce electric field noise \cite{kuehn2006}. At least two experiments \cite{Mamin2013,kim2015} found reduction in noise in NV-center diamonds by polymeric layering.

The complexity of surface morphology, physics and the fact that small-distance electric field noise occurs at different frequencies make a universal theory of the noise or identifying a single source of noise a challenge \cite{brownnutt2015}. Nevertheless, there is solid evidence supported by microscopic and phenomenological theories and experiments to suggest that adsorbed surface impurities are at work in generating fluctuating electric fields at the position of surface trapped ions \cite{henkel2008,turchette2000,safavi-naini2011,hite2012,brownnutt2015}. The fluctuating electric fields at the position of the ions are produced by the variations in the adsorbed impurity dipole moments, on trap electrode surfaces, due to coupling with the surface phonons \cite{safavi-naini2011,brownnutt2015}. Detailed calculations of the noise power spectra with density of impurity adsorption on gold surfaces, their  distance, temperature and frequency dependence, the effect of impurity mobility and surface crystalline orientation have helped elucidate some of the challenging aspects of noise production in ion microtraps \cite{safavi-naini2011,safavi-naini2013,kim2017,jooya2018,jooya2019}. 
When the impurity concentration is high on the trap surface, correlated dynamics can become important, when phonons emitted by one impurity can be absorbed by another.

In this work, we consider the effect of such collective couplings on the noise spectral power, where the adsorbed fluctuating dipoles conspire to emit to and absorb from the same phonon mode.
Coherent radiation has been of long-standing interest since Dicke's seminal work demonstrating the enhanced radiance among correlated emitters \cite{dicke1954,GROSS1982}. We also extend the usual two-level fluctuator paradigm to include anharmonicity of levels and study the frequency and temperature influence of anharmonic behavior of the fluctuator dipoles with the surface, on the noise power spectrum. By exploiting the symmetry properties of the phonon jump operators which determine the evolution of a Linbladian master equation for $N$ $M$-level fluctuators, we can make level-specific statements on how the noise emerges from the convolution of two-time correlators. We show that coherently radiating adatoms, when distributed in mutually independent patches whose relative frequency falls with their size, $\sim 1/N$, herald the emergence of 1/f noise.  

\section{Interaction Hamiltonian and many-adatom master equation}

\begin{figure}[ht!]
    \centering
    \includegraphics[scale=.5]{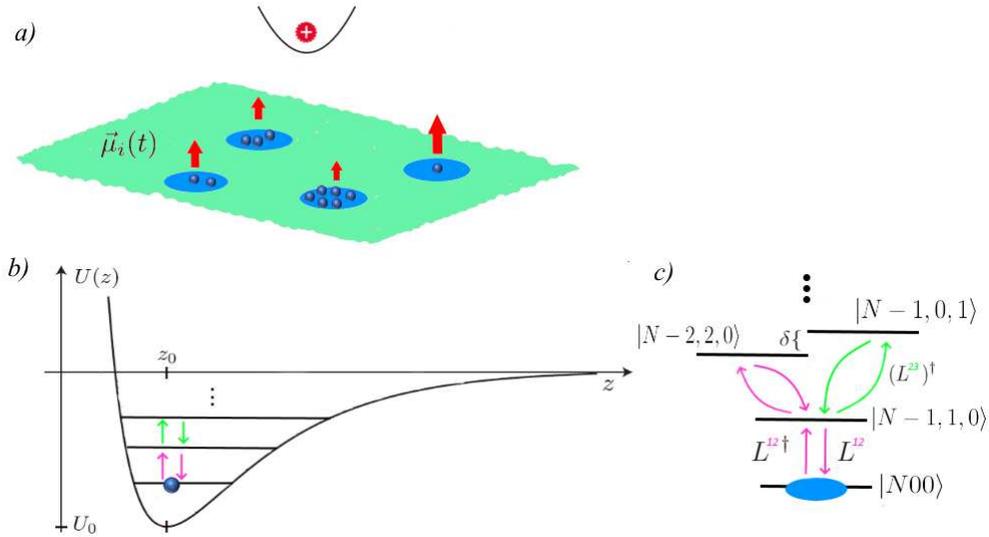}
    \caption{$a)$ An ion trapped above the surface suffers motional jitter due to the fluctuating electric fields induced by correlated dipoles on the surface.  The surface dipoles, adatom impurities, interact with the surface in the exp-3 potential, seen here in panel $b)$ and analytically represented in Eq.~\ref{eq:potential}, which supports bound vibrational levels. Collective symmetric states of the dipoles span the many-body density matrix. $c)$ Jump operators, $\hat{L}^{\mu\nu}$, describe transitions between these states which absorb or emit a phonon. An energy level diagram for a patch of $N$ adatoms illustrates the two lowest transitions. The leftmost arrows indicate the transitions between the 1$^{\text{st}}$ and 2$^{\text{nd}}$ levels while the rightmost indicate the transitions between the 1$^{\text{st}}$ and 3$^{\text{rd}}$ level. The shift $\delta$ in energy is due to the anharmonicity of the interaction.}
    \label{fig:modelsystem}
\end{figure}

The primary source of noise in ion microtraps originates from impurity atoms that are adsorbed on metallic electrode surfaces. At large atom-surface separations, the potential is determined by the atomic dynamical polarization and is attractive, while repulsive forces dominate at short distances, where the adatom wave function overlaps with the surface. This general binding is captured by the so-called exp-3 \cite{Hoinkes1980} potential, an extension upon the eponymous Buckingham potential \cite{buckingham1938}
\begin{equation}\label{eq:potential}
U(z)=\frac{\tilde{\beta}}{\tilde{\beta}-3} U_{0}\left[\frac{3}{\tilde{\beta}} e^{\tilde{\beta}\left(1-z / z_{0}\right)}-\left(\frac{z_{0}}{z}\right)^{3}\right],
\end{equation}
where  $U_0$ is the potential depth at the equilibrium position $z_0$. $\tilde{\beta}=\beta_0z_0$ where $\beta_0$ is the reciprocal range of repulsion and $z$ is the distance of the adatom from the surface as seen in Fig.~\ref{fig:modelsystem}. The potential can also depend on the transverse ($x$,$y$) coordinates due to surface roughness or impurity mobility on the surface. Here, we take these contributions to average out or otherwise be weak as would be expected on a metal surface where the electrons are smeared out \cite{safavi-naini2011}. Thus, we only concern ourselves with motion normal to the surface.
{Bound levels, supported in the potential, see Fig.~\ref{fig:modelsystem}, possess permanent dipole moments, whose average magnitude can be approximated as
$d_{\mu}\equiv0.47 e a_{0}^{1 / 2} \alpha^{3 / 2}\langle \mu| {z^{-4}}| \mu\rangle$ where $a_0$ is the Bohr radius and $\alpha$ is the atomic polarizability \cite{antoniewicz1974,safavi-naini2011}. Transitions between levels are possible by absorbing or emitting phonons. 

Near $z_0$ the harmonic approximation holds and the fundamental overtone transition frequency can be written as 
\begin{equation}\label{eq:omeganaught}
\omega_{0}=\omega_{12}\approx \sqrt{\frac{3U_{0} \left(\tilde{\beta}^{2}-4 \tilde{\beta}\right)}{m z_{0}^{2}(\tilde{\beta}-3)}},
\end{equation} 
where $m$ is the adatom mass, indicating that heavy impurities lead to lower frequency phonon absorption or emission \cite{safavi-naini2011}. A consequential feature of this interaction potential, c.f. Fig.~\ref{fig:modelsystem}, is that for increasing mode coupling to the surface phonons, anharmonic behavior will have observable influence on the noise power spectrum.
}

The displacement $\mathbf{u}_i$ of an atom in the lattice from its equilibrium position $\mathbf{r}_i^0$ is described in terms of the creation operators of the phonon eigenmodes $a_k$ by $\mathbf{u}_{i}=\sum_{k} \sqrt{\frac{h}{2 \text{N} \mathcal M \omega_{k}}} \vec{\epsilon}_{\zeta}{(\mathbf{k})}\left[a_k e^{i \mathbf{k} \cdot \mathbf{r}_{i}^{0}}+\mathbf{h . c .}\right]$ where N is the number of atoms in the bulk and $\mathcal M$ is their mass. The mode index, $k\leftrightarrow(\mathbf{k},\zeta)$, describes both the quasimomentum $\mathbf{k}$ and the polarization index $\zeta$ \cite{CrisZoll1}. For each quasimomentum, the normalized vectors
$\epsilon_{\zeta}(\mathbf{k})$ describe the three orthogonal phonon polarizations ~\cite{CrisZoll,safavi-naini2011,Schfer2002}. The fluctuating positions of these bulk atoms will affect the adatoms trapped in the potential $U$ thus affecting the vibrational state of the adatom. This coupling can described to first order in $\mathbf{u}_i$ as  $U\left(z,\left\{\mathbf{r}_{i}\right\}\right) \simeq U\left(z,\left\{\mathbf{r}_{\mathbf{i}}^{0}\right\}\right)+\sum_{i} \nabla U\left(z,\left\{\mathbf{r}_{\mathbf{i}}^{0}\right\}\right) \mathbf{u}_{i}$. Note that as we previously argued, we are interested in the normal motion, so that  $\nabla\rightarrow\partial_z$. Transitions between levels not only produce changes in the adatom energy but also variations in the dipole moment. These varying dipole moments cause electric field fluctuations at the position of the ion leading to motional jitter of the ionic qubit. {The exchange of phonons has been treated theoretically for isolated adatoms, where the emission of a phonon from one adatom does not affect the remaining adatoms \cite{safavi-naini2011}. 
Here, we expand upon this description and incorporate the collective interaction of adatoms with the phonon modes.
}

We consider a system of $N$ identical atomic impurities interacting with the surface through $M$ bound states, whose density matrix is denoted by $\hat{\rho}$. A transition from level $\nu$ to $\mu$ ($E_\mu < E_\nu$) on atom $n$ is described by the operator $\hat{b}^{\mu \nu}_n =\ket{\mu}\bra{\nu}$ accompanied by the emission of a phonon into the bulk of frequency $\omega_{\mu\nu}=\omega_\nu-\omega_\mu$.  Following similar arguments as those found in \cite{CrisZoll} for atom-photon coupling, we derive the master equation for our system. The Hamiltonian describing the interaction of $N$ adsorbed atomic impurities with the surface can be written in the interaction picture and in the rotating-wave approximation \cite{CrisZoll},
\begin{equation}
    \hat{\mathcal{H}} = -i\hbar \sum_n^N\sum_{\mu<\nu}^M\sum_{k}\left[g^{\mu\nu}_n(t,k)a_kb^{\mu\nu\text{ } \dagger}_{n}+ g_n^{\mu \nu *}(t,k)a^\dagger_kb^{\mu\nu}_{n}\right],
\end{equation}
where the atom-phonon coupling term is 
\cite{CrisZoll} 
\begin{equation}\label{eq:coupling}
g_n^{\mu \nu}(t,k) = \sqrt{\frac{\hbar}{2\text{N}{\mathcal{M}\omega_{k}}}} \epsilon_\zeta({\vec k})\cdot {\nabla U} e^{-i(\omega_{\vec{k}}-\omega_{\mu\nu})t+i\vec{k}\cdot \vec{r}_n}
\end{equation}
and the relation $\mu<\nu$  from now on will be dropped from the notation and implicitly assumed.
To arrive at a description of the adatoms alone, we formally solve the phonon dynamics by evaluating the commutator of the noise operators 

\begin{equation*} 
\gamma_{mn }^{\mu\nu}(t-t^{\prime})=\sum_{k,k'}[g^{\mu\nu}_n(t) a_k,g_m^{\mu\nu *}a_{k'}^{\dagger}(t')]=\sum_{m\neq n, k}\left|\kappa_{k}\right|^{2} \mathrm{e}^{-i\left(\omega_{k}-\omega_{\mu\nu}\right)(t-t^{\prime})+i \boldsymbol{k} \cdot\left(\boldsymbol{x}_{m}-\boldsymbol{x}_{n}\right)}
\end{equation*}
where $\kappa_k$ is the non exponential part of Eq.~\ref{eq:coupling} and $m\neq n$ will be dropped from the notation. Following a corresponding derivation for atoms coupled to photons in free space \cite{CrisZoll,RJThesis}, this commutator is central in developing an effective model for the adatom dynamics.  We first perform the sum over the polarization directions, $\zeta$, which leaves only a dependence on $\mathbf{k}<\mathbf{k}_c$ \cite{CrisZoll} where $\mathbf{k}_c$ is the Debye cutoff of the lattice. The system dynamics are then fully captured by the cooperative decay $\Gamma_{mn}^{\mu\nu}$ as well as the induced energy shift $\delta \omega_{mn}^{\mu \nu}$ \cite{CrisZoll} which in analogy to atomic systems can be be extracted from
\begin{equation}\label{eq:gammaINT}
\int_{0}^{\Delta t} d t_{1} \int_{0}^{t_{1}} d t_{2} \gamma_{mn}^{\mu\nu}\left(t_{1}-t_{2}\right) \equiv\left(\frac{1}{2} \Gamma_{mn}^{\mu\nu}+\mathrm{i} \delta \omega_{mn}^{\mu\nu}\right) \Delta t.
\end{equation}
We assume the usual Markovian approximation that bath correlations decay much faster than the correlation among the adatoms and that retardation effects can be ignored \cite{Manzoni_2018}. The latter approximation holds when the adatom separation is much less than the phonon correlation length, i.e. $r_{mn}\ll\frac{c}{\Gamma_0}$ where  $\Gamma_0$ is a characteristic  phonon spontaneous emission rate and $r_{mn}$ is the interatomic distance \cite{LehmbergerI,LehmbergerII}. Under these conditions, we can solve for the cooperative decay $\Gamma^{\mu\nu}_{mn}$ and energy shift $\delta\omega_{m,n}$ 
and obtain an effective description for the correlated adatom ensemble, which we write in Lindbladian form
\begin{equation}
\partial_t \hat{\rho} = \sum_{\mu \nu}^M i[\hat{\rho},\hat{\mathcal{H}}^{\mu\nu}] + \mathcal{L}^{\mu\nu}(\hat{\rho})
\end{equation}
with 
\begin{eqnarray}
 \label{ndml}
 \hat{\mathcal{H}}^{\mu\nu} = \sum_{mn}^N \hat{b}_{m}^{\mu\nu\dagger}\hat{b}_{n}^{\mu\nu}( \omega_{\mu \nu} +\delta\omega_{mn}^{\mu\nu})\\
 \mathcal{L}(\hat{\rho}) = \sum_{mn}^N\frac{\Gamma _ { m  n }^{ \mu \nu }}{1-e^{-\beta\omega_{\mu\nu}}}\left(\hat{b} _ { m }^{ \mu \nu } \hat{\rho} \hat{b}_n^{ \mu \nu \dagger}-\frac{1}{2}\left\{ \hat{b}_n^{ \mu \nu \dagger}\hat{b}_m^{ \mu \nu },\hat{\rho}\right\}+\right. \\ \nonumber
 \left. e^{-\beta \omega_{\mu \nu}} \left(\hat{b} _ { m }^{ \mu \nu \dagger} \hat{\rho} \hat{b}_n^{ \mu \nu }-\frac{1}{2}\left\{ \hat{b}_n^{ \mu \nu } \hat{b}_m^{ \mu \nu \dagger},\hat{\rho}\right\}\right)\right).
\end{eqnarray}
The Hamiltonian describes that a phonon after being emitted from an adatom transitioning from state $\nu$ to state $\mu$ can be absorbed by one of the other adatoms, promoting it from state $\mu$ to state $\nu$. This is analogous to the exchange of virtual photons in dense multi-level atomic gases. The dissipative interaction of adatoms, $m$ and $n$, in state, $\mu$ and $\nu$, is given by the matrix $\Gamma _ { m n } ^ { \mu \nu }=\frac { 3 } { 2 } \Gamma_0 ^ { \mu \nu } f \left( k ^ { \mu \nu } r_{ m n } \right)$ where the level transition frequency is $\Gamma_0^{\mu\nu}=\frac{|\langle\mu|\nabla U|\nu\rangle|^{ 2 }\omega _ { \mu \nu } } { 2 \pi \hbar c ^ { 3 } \mathscr{N}}$ given that $k^{\mu\nu}= \frac{ \omega _ { \mu \nu } } { c } $ and $\mathscr{N}$ is the bulk density. Both coherent exchange process and the decay carry an intrinsic dependence on the distance of any pair of two adatoms, given by the function $f$, that can be obtained from the explicit solution of the integral in Eq.~\ref{eq:gammaINT} \cite{CrisZoll}.
The inverse temperature is $\beta=\frac{\hbar}{k_b \text{T}}$, with T as the temperature, but the quantity of merit in our derivations is $\beta\omega_0$, where $\omega_0$ is the main transition frequency.

The interaction potential is harmonic about $z_0$, and as such the fundamental transition rate  can be derived,  $\Gamma_0^{12}=\Gamma_{0} \approx \frac{1}{4 \pi} \frac{\omega_{0}^{4} m}{c^{3} \mathscr{N}}$. This expression can be useful when we return to the $N>1$ case, as in the long-wavelength limit, where $k^{\mu \nu}r_{mn}\ll 1$, the decay is spatially independent such that $\Gamma _ { m n } ^ { \mu \nu }$ becomes constant with elements, $\Gamma_0^{\mu \nu}$. 
{In this limit, an emitted virtual phonon is equally likely to be  absorbed by any of the adatoms, revealing that the system is fully symmetric under exchange of two adatoms. All such transitions are captured by a collective operator, $\hat{L}^{\mu \nu} = \sum_{i}^N \hat{b}_i^{\mu \nu}$, acting on all adatoms equally. This affords a particularly simple description, where we can}
naturally reduce Eq.~\ref{ndml} to 
\begin{eqnarray}\label{jumprho}
\partial_t \hat{\rho}= \sum _ { \mu \nu }^M -\frac{i}{h}[ H_{\mu \nu}  \hat{L}^{\mu \nu \dagger } \hat{L}^{\mu \nu}  , \hat{\rho}]+ 
\frac{\Gamma_0 ^{\mu \nu}}{1-e^{-\beta \omega_{\mu\nu}}}\times \\ \nonumber  \left( \hat{L} ^{\mu \nu} \hat{\rho} \hat{L}^{\mu \nu \dagger } - \frac { 1 } { 2 } \left\{ \hat{L} ^{\mu \nu \dagger } \hat{L}^{\mu \nu}, \hat{\rho} \right\}+ 
e^{-\beta \omega_{\mu\nu}}\left( \hat{L} ^{\mu \nu \dagger } \hat{\rho} \hat{L} ^{\mu \nu} - \frac { 1 } { 2 } \left\{ \hat{L} ^{\mu \nu} \hat{L} ^{\mu \nu \dagger }  , \hat{\rho} \right\} \right)\right).
\end{eqnarray}
The action of the fully symmetric operators strictly isolates the symmetric subspace from all other states \cite{molmer2018,Geremia2010}. We denote the symmetric states as $\ket{m_1 m_2...m_M}$ where $m_i$ etc indicates the number of atoms in level $i$ with $N=\sum_i m_i$. 
Now, the action of $L_{\mu \nu}$ on a state $\ket{m_{\mu}m_{\nu}}$ is given by 
\begin{equation}
 \hat{L}_{\mu \nu}\ket{...,m_{\mu},...,m_{\nu},...}=\sqrt{(m_{\mu}+1)m_{\nu}}\ket{...,m_{\mu}+1,...,m_{\nu}-1,...}.
\end{equation}
Furthermore, Eq.~(\ref{jumprho}) decouples the evolution of populations in the symmetric basis from coherences. Assuming an unexcited initial state, the system remains in the subspace spanned by ${N+M-1 \choose N}$ populations of the symmetric subspace, denoted by $\rho_{m_1,...,m_M}$. These populations are eigenstates of the Hamiltonian, which therefore does not contribute to the system evolution. This renders an effective classical model for the interacting adatoms.

This is understood directly by writing Eq.~\ref{jumprho} in the symmetric population basis
\begin{eqnarray}\label{directM}
\dot{\rho}_{m_1,...,m_{M}}=\sum_{\mu,\nu} F^{\mu\nu}(m_{\nu} \left(m_{\mu}+1\right)\left(e^{-\beta\omega_{\mu\nu}}\rho_{...,m_{\mu}+1,...,m_{\nu}-1,...}-\rho_{m_1,...,m_M}\right)\\+ \nonumber
m_{\mu}\left(m_{\nu}+1\right)\left(\rho_{...,m_{\mu}-1,...,m_{\nu}+1,....}-e^{-\beta\omega_{\mu\nu}}\rho_{m_1,...,m_M}\right))
\end{eqnarray}
where $F^{\mu\nu}=\frac{\Gamma_0^{\mu\nu}}{1-e^{-\beta\omega_{\mu\nu}}}$. {Eq.~\ref{directM} defines a linear matrix equation for the density matrix components, $\dot{\rho}_i =\sum_jM_{ij}\rho_j $} where $\rho_j$ is the $j^{\text{th}}$ state in the symmetric basis we have described and $M_{ij}$ is the transition element between two of the states. 

Now that we have a description for the dynamics of the vibrational states of our system, we can write an equation for the change in dipole moment. Choosing $\hat\rho_i=\ket{i}\bra{i}$ as the projector of the symmetric state $i$ we can write the dipole operator of this state as $\hat{\mu}(t)=\sum_iD_i\hat{\rho_i}(t)$, where $D_i$ is the associated dipole moment magnitude. This value is given by the sum over the dipole moments of the populated levels of the $i^{\text{th}}$ state weighted by their occupation; for instance the dipole moment of a state $\rho_5=\ket{201}$ is given by $D_5=2d_1+d_3$. We can proceed to calculate the induced electrical field noise of these fluctuators to which an ion located at close proximity from the trap surface would be sensitive.

\section{Noise spectrum of correlated anharmonic fluctuators} 
We calculate the magnitude of the electric field noise by invoking the Wiener–Khinchin theorem \cite{Wiener1930,Khintchine1934} which in this context states that the electric field noise power spectrum is the Fourier transform of the autocorrelation function of $\hat{\mu}$, 
\begin{equation}
    S(\omega) = \int_{-\infty}^{\infty} d\tau \left[ \langle \hat{\mu}(\tau)\hat{\mu}(0) \rangle -\langle \hat{\mu}(0) \rangle^2 \right] e^{i\omega \tau} \label{eq:noisy}.
\end{equation}
The correlation functions can be evaluated using the quantum regression theorem \cite{CrisZoll1}.
The first term in Eq.~\ref{eq:noisy} can be written as
\begin{equation*}
    \langle \hat{\mu}(\tau)\hat{\mu}(0) \rangle = \sum_{i,j} D_i D_j \langle \hat{\rho}_i(\tau)\hat{\rho}_j(0) \rangle .
\end{equation*}
Since the operators $\rho_i$ just represent projections, they satisfy the same equation of motion of the density matrix $\partial_t \langle \hat{\rho}_i(\tau)\hat{\rho}_j(0) \rangle = \sum_{j'} M_{ij'} \langle \hat{\rho}_{j'}(\tau)\hat{\rho}_j(0) \rangle$.
As the matrix is diagonalizable, $M = A^{-1}\mathcal{D}A$ with eigenvalues $\{\lambda_k \leq 0\}$, a particularly simple solution is in terms of the eigenmodes
\begin{equation*}
    \langle \hat{\rho}_i(\tau)\hat{\rho}_j(0) \rangle = \sum_k A_{ik}^{-1}\sum_l A_{kl} \langle \hat{\rho}_l(\tau)\hat{\rho}_j(0) \rangle e^{\lambda_k \tau}.
\end{equation*}
The first term in Eq.~\ref{eq:noisy} can, thus, be written as
\begin{equation*}
    \langle \hat{\mu}(\tau)\hat{\mu}(0) \rangle = \sum_{i,j} D_i D_j \sum_k A_{ik}^{-1}\sum_l A_{kl} \langle \hat{\rho}_l(0)\hat{\rho}_j(0) \rangle e^{\lambda_k \tau}.
\end{equation*}
Note that the term $\lambda_k=0$ corresponds to the steady state and can be simplified into $\langle \hat{\mu}(0)\hat{\mu}(0) \rangle$, cancelling the second term in Eq.~\ref{eq:noisy} as expected. We further note that the projector can be simplified into the steady-state value
\begin{equation*}
    \langle \hat{\rho}_l(0)\hat{\rho}_j(0) \rangle = \langle \hat{\rho}_j(0) \rangle \delta_{lj} = \rho_j(0)^\text{ss} \delta_{lj}.
\end{equation*}
The Fourier integral can now be performed analytically, when observing the time symmetry of the correlator $\langle \hat{\mu}(\tau)\hat{\mu}(0) \rangle = \langle \hat{\mu}(-\tau)\hat{\mu}(0) \rangle$ in the steady state
\begin{eqnarray*}
S(\omega) &= \sum_{i,j} D_i D_j \sum_{k \setminus \{\lambda_k=0\}} A_{ik}^{-1} A_{kj} \rho_j(0)^\text{ss} \int_{-\infty}^{\infty} d\tau e^{\lambda_k |\tau|}e^{i\omega \tau} \\
&= -\sum_{i,j} D_i D_j \sum_{k \setminus \{\lambda_k=0\}} A_{ik}^{-1} A_{kj} \rho_j(0)^\text{ss} \frac{2 \lambda_k}{\lambda_k^2 + \omega^2},
\end{eqnarray*}
where the integral converges because $\lambda_k < 0$. For the same reason, the noise spectrum is strictly positive definite. A concise interpretation can be obtained by defining 
\begin{equation*}\label{eq:coeffNoiseSpec}
    C_k = 2 \sum_{i,j} D_i D_j A_{ik}^{-1} A_{kj} \rho_j(0)^\text{ss},
\end{equation*}
such that the noise spectrum becomes a sum of Lorentzians 
\begin{equation}\label{eq:NoiseSpec}
    S(\omega) = \sum_{k \setminus \{\lambda_k=0\}} C_k \frac{ -\lambda_k}{\lambda_k^2 + \omega^2}.
\end{equation}
This simple exact solution expresses the remarkable equivalence of the electric field noise from an $N$-body correlated system to that of $N+M-1\choose N$-1 (the number of states excluding the steady state) {independent effective two-level fluctuators. In this picture, each fluctuator has a weight $C_k$ with an effective lifetime $\tau_k=\lambda_k^{-1}$. This provides a natural order of decay that can be useful for the simplified analysis of noise spectra at high and low frequencies.}

\section{Correlated noise discussions}
The electric field noise emanating from surface impurity fluctuations has a complicated dependence on crucial parameters in the experiments, but can be generally parameterized as $S(\omega) \propto T^{\alpha_1} H^{\alpha_2} \omega^{\alpha_3}$, where $H$ is the distance from the surface to the qubit and the exponents are some constants. The complex adatom-surface interaction is modelled in Eq.~\ref{eq:potential}, with transition energies $\omega_{\mu\nu}$, dipole moments, $d_{\mu}$, transition rate constants $\Gamma_{\mu\nu}$, and temperature $T$. Numerical calculations indicate that $d_\mu$ decreases linearly with increasing vibrational level while $\Gamma_{\mu(\mu+1)}$ peaks at some intermediate level \cite{safavi-naini2011}.

\begin{figure}[h]
    \centering
    \includegraphics[scale=.8]{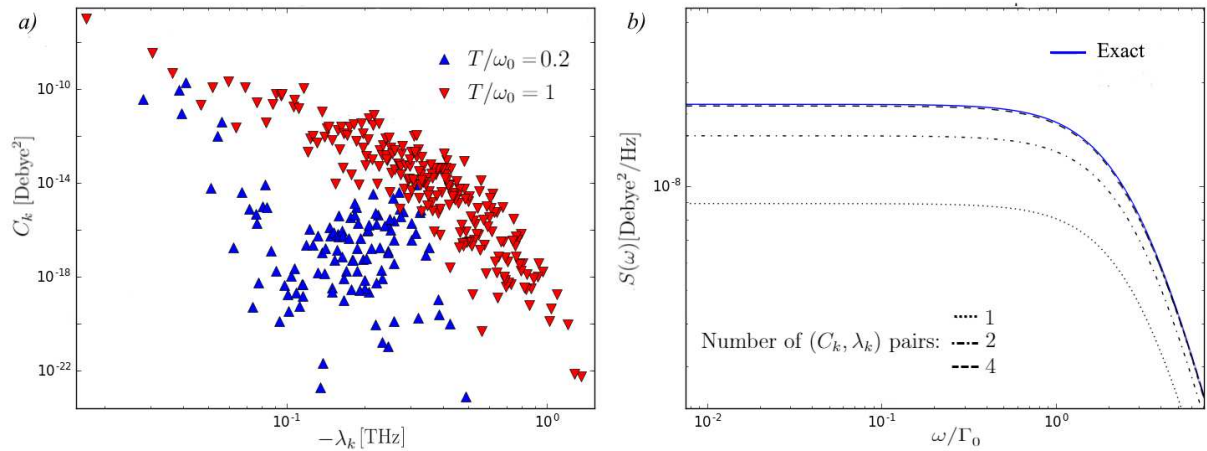}
    \caption{$a)$ The coefficients $C_k$ in Eq.~\ref{eq:NoiseSpec} are shown as a function of eigenvalues, $\lambda_k$ for $N=3$. The up (blue) and down (red) pointing triangles correspond to numerically exact calculations at  $T/\omega_0=0.2, 1$, respectively. The widespread distributions of weights $C_k$ over 10 orders of magnitude suggest an approximation scheme utilizing the largest $C_k$. $b)$ The noise power spectrum illustrates the point:  with only the highest weighted ($C_k$, $\lambda_k$) pairs, here up to four pairs, results converge to the exact values for $T/\omega_0=0.2$. The convergence will only be more rapid for higher temperatures. For all calculations $\beta_0=1.86$\AA$^{-1}$ and $z_0=3.1$\AA.}
    \label{fig:CkvsLk}
\end{figure}

The noise power spectra are shown in Fig.~\ref{fig:CkvsLk}(b) for two typical low temperatures. Fig.~\ref{fig:CkvsLk}(a) displays the decomposition of terms in Eq.~\ref{eq:NoiseSpec}. Each component $\lambda$ refers to one Lorentzian. While it is evident that many Lorentzians contribute at both low and high temperatures, only a few $\lambda$ terms dominate the noise power spectrum decomposition.

The effect on the noise power spectrum is shown in Fig.~\ref{fig:CkvsLk}(b), where the sum of only a few $\lambda$ terms are necessary to reproduce the "exact" numerical results (calculations converge with $M=10$), reaffirming the expectation that level population is thermally activated and determined by the Boltzmann factor, $\mathcal{T} = e^{-\beta \omega_{0}}$.

\begin{figure}[ht!]
    \includegraphics[scale=.45]{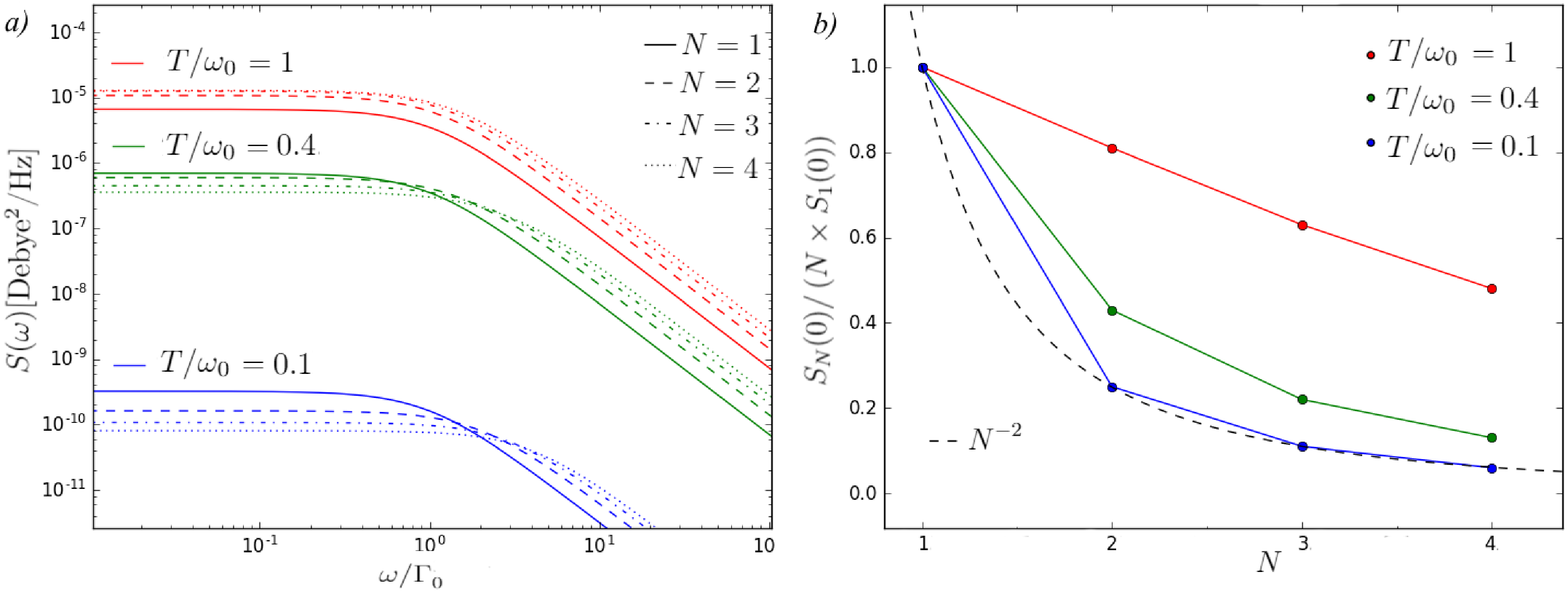}
    \caption{The noise spectral function for dipoles with $m=100$ amu, interacting with the surface potential at depth $U_0=250$ meV at  $T/\omega_0=0.1,0.4$, and $1$. $a)$ The correlated noise spectrum is shown for different patch sizes. At $T/\omega_0 <1$, correlated superradiant decay results in a lowering of the white noise with increasing number of correlated adatoms. Thus, in this regime one adatom is a larger source of noise than 2 or more correlated adatoms. With $T/\omega_0\sim 1$, superradiant decay does not prevent an increasing noise magnitude as a function of increasing $N$. $b)$ The white noise spectra per adatom at different $T/\omega_0$ is shown to decrease with increasing $N$ where in the low temperature limit the $N^{-2}$ scaling appears as predicted by Eq.~\ref{eq:2LLT}.}
    \label{fig:combinedevolve}
\end{figure}

{At low temperatures $T/\omega_0 \ll 1$, the dominant transitions occur near the harmonic minimum, $z_0$, such that only the lowest two states are important. The noise spectrum can be expanded perturbatively in powers of $\mathcal{T}$,
}
\begin{equation}\label{eq:2LLT}
S_N(\omega)=2(d_1-d_2)^{2}\frac{N\Gamma_0}{(N\Gamma_0 )^2+\omega^2} \mathcal{T} + \mathcal{O}(\mathcal{T}^2)
\end{equation}
An immediate observation, in comparing to Eq.~\ref{eq:NoiseSpec}, is $\lambda=N\Gamma_0$  and its linearity in the number of adatoms $N$. The proportionality of the eigenvalues with $N$ is exploited later to describe how the ubiquitious $1/f$ noise emerges. The equivalent coefficients, $C_k$, are independent of $N$.  The frequency independent white noise per adatom, $\frac{1}{N}S_N(\omega=0)$, now scales as $\frac{1}{N^2}$, as observed in Fig.~\ref{fig:combinedevolve}(b). This suggests that at low temperatures, correlated phonon transitions conspire to suppress electric field noise, see Fig.~\ref{fig:combinedevolve}(a). In this regime, a larger number of adatoms not only decreases the noise per adatom but, in fact, decreases the total noise of $N$ adatoms to a value lower than that of a single adatom ($N=1$). This is a reasonably {\it counterintuitve and surprising} consequence of superradiant decay, as applied to noisy adatoms on surfaces. 

{\it Superradiant decay:} {Fig.~\ref{fig:combinedevolve}a displays the noise spectrum for three different temperatures. We recover the superradiant behavior at low temperatures as discussed above, where the overall low-frequency noise is attenuated. This effect is weakened at intermediate temperatures and eventually reversed, and the total noise increases with increasing $N$. Remarkably, the noise per adatom generally remains suppressed as seen in Fig.~\ref{fig:combinedevolve}b. At higher temperatures, the influence of interaction potential anharmonicity
on the level structure and transition rate constants -- see below -- increases. As analyzed in Fig.~\ref{fig:CkvsLk}, the contribution of different $(\lambda_k, C_k)$ sets to Eq.~\ref{eq:2LLT} shifts such that at $\omega/(N\Gamma_0)\approx1$, the spectral dependence turns over from constant to $1/f^2$.  Throughout, we ensure that the adatom remains bound to the surface and cannot be thermally released.} 

\begin{figure}[ht!]
    \centering
    \includegraphics[scale=1.7]{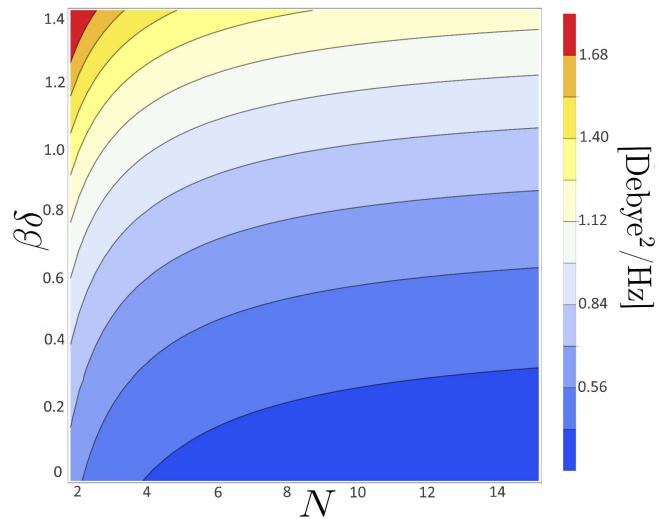}
    \caption{The coefficient of $\mathcal{T}^2$, $C_2(N,\delta)$ in Eq.~\ref{eq:2ONS}, is shown as a function of $N$ and $\beta\delta$ for scaled quantities $d_{1,2,3}=1,.8,.6$  and $\Gamma^{(1,2)(2,3)}_0=1,1.4$ respectively. For a fixed $\beta\delta$ the many-body contribution from $C_2(N,\delta)$ to the total white noise spectrum plateaus at large $N$ with the plateau height increasing with increasing anharmonicity.}
    \label{fig:Anharmonicity}
\end{figure}

{\it Anharmonicity:} {The anharmonic nature of the interaction potential, Eq.~\ref{eq:potential}, influences the dynamics. This influence grows with increasing thermal fluctuations. At $T/ \omega_0=0.2$, the dominant effective two-level fluctuator approximates the exact white noise spectrum to within a few percent. This discrepancy grows quickly with increasing temperature. The most immediate extension is to consider three levels instead of two. This introduces anharmonic contributions to the noise power spectrum which come from the higher order terms in the expansion in $\mathcal{T}$ in Eq.~\ref{eq:2LLT}. A convenient parametric measure of the anharmonicity is found by expanding $U(z)$ in Eq.~\ref{eq:potential} about $z_0$ to the third order in $z$, such that the frequency asymmetry between level spacing is,
\begin{equation}
\delta=\omega_{0}-\omega_{23}\approx\frac{5\hbar}{24
m}\frac{(\tilde{\beta}^2-20)^2}{z_0^2 (\tilde{\beta}-4)^2}.
\end{equation}
In Fig.~\ref{fig:Anharmonicity}, we investigate how increasing anharmonicity, $\delta$, affects the noise spectrum of $N$ correlated fluctuators. With increasing $N$, the spectral density plateaus, but with increasing $\delta$, the plateau heights also increase. 
}
At higher temperatures, higher orders in $\mathcal{T}$ are necessary and more anharmonic transitions are needed. Assuming only the next neighbor transitions, and expanding to second order in $\mathcal{T}$, one finds $S_N(0)=C_1(N)+C_2(N,\delta)\mathcal{T}^2$ with
\begin{equation}\label{eq:2ONS}
C_2(N,\delta)=2\frac{(d_1-d_2)^{2}(3N-1)}{(N-1)N\Gamma^{12}_0}+
2(d_1-d_3)\left(\frac{2(d_1-d_2)}{N\Gamma^{12}_0}+\frac{d_1-d_3}{\Gamma^{23}_0}\right)e^{\beta \delta}
\end{equation}
Here we leave the ($\mu\nu$) indices (1,2) and (2,3) to make transparent the decay matrix elements. 

{\it $1/f$ noise emergence:} The overall noise affecting an ion above a trap surface originates from a mesoscopic area of randomly scattered impurities with differing local densities which together define a surface coverage. The interaction with the phonon hub is, therefore, given by the separation of all adatom pairs. Instead of reintroducing this spatial dependence, we provide a simple and insightful approximation of this complex system by segregating the surface into patches, whose adatoms are fully correlated, while each patch interacts independently with the phonon hub (Fig.~\ref{fig:modelsystem}). Thus, the system noise can be approximated as a weighted sum of the patches of size $N$ 
\begin{equation}
\label{eq:total}
S_{tot}(\omega)=\sum_{N=1}^{N_{\rm max}}\mathcal{D}(N)S_N(\omega)
\end{equation}
with a corresponding relative weight factor $\mathcal{D}(N)$.
In particular, at low temperatures the noise spectrum can be written as $S_N(\omega)=\mathcal{C}\frac{\Gamma_0 N}{(\Gamma_0 N)^2+\omega^2}$ where $\mathcal{C}$ are the coefficients of the Lorentzian in Eq.~\ref{eq:2LLT}. Since $\mathcal{C}$ has no $N$ dependence, it can be moved out of the sum and normalized. 

\begin{figure}[ht!]
    \centering
    \includegraphics[scale=.5]{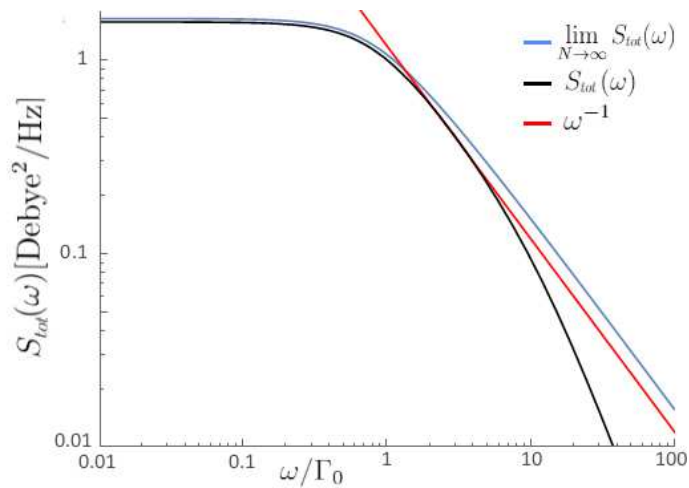}
    \caption{ The emergence of $1/f$ noise with a $\mathcal{D}(N) =\frac{1}{N}$ patch distribution. The black curve results from a sum over $N_{\rm max}=10$ correlated adatoms in Eq.~\ref{eq:total} and the blue curve is the $N_{\rm max}\rightarrow\infty$ limit in  Eq.~\ref{eq:sumpink}. With increasing number of correlated adatoms in each patch, the noise spectrum turns from $1/f^2$ to $1/f$.}
    \label{fig:sumpink}
\end{figure}

Next, we consider a distribution whereby each patch is weighted by  $\mathcal{D}(N)=N^{-1}$. Then
\begin{equation}
\label{eq:sumpink}
\lim_{N_{\rm max}\rightarrow\infty}
S_\text{tot}(\omega)=\frac{\mathcal{C}}{2}\left(-\frac{\Gamma_0}{\omega^2}+\frac{\pi \coth\left(\frac{\pi \omega}{\Gamma_0}\right)}{\omega}\right).
\end{equation}
At small frequencies, $\omega/\Gamma_0 \ll 1$, the noise thus approaches a constant value, $\pi^2/(6\Gamma_0)$, while $1/f$ behavior emerges for $\omega/\Gamma_0>1$, see Fig.~\ref{fig:sumpink}. For finite sums, the spectral behavior shows  $1/f$ noise across a finite frequency range, before returning to the typical $1/f^2$ behavior at the highest frequencies. This derivation forms an interesting complement to the phenomenological derivation of $1/f$ noise by Dutta and Horn \cite{dutta1981}. There, a distribution of  oscillators is assumed whose characteristic time scales $\tau$ are distributed as $\mathcal{D}(\tau) \propto \tau^{-1}$. Here, each patch is demonstrated to have a characteristic timescale $\tau_N \sim \frac{1}{\Gamma_0 N}$ due to superradiant decay: Larger patches have faster dynamics and therefore shorter decay timescales. Thus, we provide a physical derivation for the ubiquitous $1/f$ noise that is independent of apriori assumptions and emerges naturally from the coupled dynamics of identical fluctuators.
At higher $T/\omega_0$, level multiplicity and  interaction anharmonicity will enter, as discussed above. We have established with exact numerical calculations that $\lambda$ vs. $N$ remains linear and the slope decreases with temperature. We have also established numerically that while the Lorentzian coefficient does have $N$ dependence at higher temperatures, $C_2(N,\delta)$  plateaus with  $N$, see Fig.~\ref{fig:Anharmonicity}. Therefore, as with the low temperature case, there is a linear region of $\lambda$ with $N$, and $\mathcal{C}$ constant. Thus while the exact shape of the noise spectrum as well as the extension of the $1/f$ noise region may vary, a $1/f$ regime could also be present for higher temperatures.

{\it Operating parameters:} 
Whether correlated adatom dynamics are expected in experiment depends largely on the adatom species, via its mass ($m$) and its specific surface interaction potential ($U(z)$), as well as the adatom density $\mathcal{R}$. Correlated dynamics will generally emerge when the dominant phonon wavelength exceeds the typically adatom-adatom separation, i.e. $\mathscr{C} \equiv \sqrt{\mathcal{R}}(2\pi c/\omega_0) \gtrsim 1$ where $c$ is the phonon speed in the surface material. Such conditions tend to be satisfied in ion traps with a large adatom concentration and relatively massive adatoms, whose vibrational energies, $\hbar\omega_0$ are low. These energies vary widely \cite{Santiago2014} from $1$ eV for hydrogen \cite{Hoinkes1980,safavi-naini2011} to $1$ meV for loosely bound heavy adatoms. For example, argon or xenon on a gold surface ($c\approx3240$ m/s) have vibrational energies $\hbar \omega_0\sim 5$ meV \cite{Ossicini1986,Stampfl2008}, such that $\mathscr{C}>1$ for moderate densities $\mathcal{R}>10^{-3}$ \AA$^{-2}$. The superradiant suppression of electric field noise is expected in such systems when the temperatures are low $T<50$ K, i.e. $\beta \omega_0 \gg 1$. More reactive adatoms such as carbon, nitrogen or oxygen, typical contaminant species, can possess higher binding energies and thus $\hbar \omega_0\gtrsim 10$ meV therefore reaching the correlated regime at greater densities, $\mathcal{R} \sim 5 \cdot 10^{-3}$ \AA$^{-2}$. On the other hand, one can satisfy the low-temperature conditions at elevated temperatures of $T \sim 100$ K while still maintaining $\mathscr{C}\gtrsim1$ with these small adatom species. Correlated dynamics should thus be at work in a wide range of current experimental ion-trap systems where we offer a microscopic derivation of the emergence $1/f$ noise at cryogenic temperatures. However, we emphasize that our theoretical framework makes no assumption on the temperatures and can thus generally be applied.

\section{Concluding remarks and outlook}
Superradiant decay, caused by the collective emission into and absorption from a phonon bath, is shown to be responsible for the suppression of electric field noise emanating from $N$ fluctuating impurities on surfaces. The electric field noise of this $N$-body correlated system maps to $N+M-1\choose N$-1 independent effective two-level fluctuators. This is facilitated by exploiting the inherent permutation symmetry of adatoms at close distances. We further find that the correlated dynamics of $N$ adatoms in surface patches with relative frequencies  $\frac{1}{N}$ can result in the ubiquitous $1/f$ noise spectrum. 

Future extensions of this correlated model for noise suppression can account for impurity diffusion and exchange of adatoms between patches. Care is then necessary to allow for spatial dependence of the phonon dynamics and the possible population of states without perfect permutation symmetry. This more complex scenario may allow the emergence of robust $1/f$ noise for a wider class of particle distributions than considered in the present article.  

Another promising direction lies in the regime of high temperatures. Interesting features at these temperatures can already be seen in Fig.~\ref{fig:CkvsLk}, where the equivalent two-level fluctuators are distributed in a more regular fashion that at low temperatures. Moreover, a single Lorentzian tends to dominate more with increasing temperature, requiring fewer pairings of ($C_k$, $\lambda_k$) for a converged description of the noise spectrum. This suggests a possible way in which the otherwise more complex computations at higher temperature calculations could be simplified. However, a comprehensive model needs to account for the possible thermal ejection of adatoms at high temperatures.

\section{Acknowledgements}
We acknowledge several helpful discussions with Arghavan Safavi-Naini and use of her notes in the early stages of this work. P. N. Thomas Lloyd thanks ITAMP for the opportunity to visit in early 2020 where these calculations began, and a virtual summer internship. This work has been supported by the
NSF through a grant for the Institute for Theoretical
Atomic, Molecular, and Optical Physics at Harvard University and the Smithsonian Astrophysical Observatory.

\section*{References}
\bibliographystyle{iopart-num}
\bibliography{1overf.bib}

\end{document}